\documentclass[aps,prl,reprint, amsmath, amssymb,superscriptaddress,nofootinbib]{revtex4-1}
\usepackage{bm}
\usepackage[utf8]{inputenc}
\usepackage[T1]{fontenc}
\usepackage[retainorgcmds]{IEEEtrantools}
\usepackage{graphicx}
\usepackage{mathrsfs}
\usepackage{amsmath}
\usepackage{indentfirst}
\usepackage{amssymb}
\usepackage{color}
\usepackage{amsfonts}
\usepackage[framemethod=TikZ]{mdframed}
\usepackage{dsfont}
\usepackage{times,txfonts}
\usepackage{appendix}
\usepackage{nicefrac}
\usepackage{enumitem} 
\usepackage{mathtools}
\usepackage{soul}
\usepackage[colorlinks=true,linkcolor=blue,urlcolor=blue,citecolor=blue,pdfusetitle]{hyperref}
\usepackage{bbold}

\mdfdefinestyle{MyFrame}{%
    linecolor=black,
    outerlinewidth=1pt,
    roundcorner=0pt,
    innertopmargin=0.77\baselineskip,
    innerbottommargin=0.75\baselineskip,
    innerrightmargin=14pt,
    innerleftmargin=14pt,
    backgroundcolor=gray!10!white}

\DeclareMathOperator{\tr}{tr}

\newcommand{\Tr}{\rm Tr}

\newcommand{\ket}[1]{\ensuremath{\left|{#1}\right\rangle}}
\newcommand{\bra}[1]{\ensuremath{\left\langle{#1}\right |}}
\newcommand{\ketbra}[2]{\ensuremath{| #1 \rangle \langle #2 |}}

\newlength{\eqboxstorage}

\begin{document}

\title{Two-point measurement of entropy production from the outcomes of a single experiment with correlated photon pairs}

\author{ Gabriel H. Aguilar}
\affiliation{Federal University of Rio de Janeiro, Rio de Janeiro, RJ, Brazil}

\author{Tha\'is L. Silva}
\affiliation{Federal University of Rio de Janeiro, Rio de Janeiro, RJ, Brazil}

\author{Thiago E. Guimar\~aes}
\affiliation{Federal University of Rio de Janeiro, Rio de Janeiro, RJ, Brazil}

\author{Rodrigo S. Piera}
\affiliation{Federal University of Rio de Janeiro, Rio de Janeiro, RJ, Brazil}

\author{Lucas C. Céleri}
\affiliation{Institute of Physics, Federal University of Goiás, P. O. Box 131, 74001-970, Goiânia, Brazil}

\author{Gabriel T. Landi}
\email{gtlandi@gmail.com}
\affiliation{Instituto de F\'isica, Universidade de S\~ao Paulo, CEP 05314-970, S\~ao Paulo, Brazil}

\begin{abstract}
Fluctuation theorems are one of the pillars of non-equilibrium thermodynamics. Broadly speaking, they concern the statistical distribution of quantities such as heat, work or entropy production. Quantum experiments, however, usually can only assess these distributions indirectly. In this letter we provide an experimental demonstration of a quantum fluctuation theorem where the distribution of entropy production is obtained directly from the outcomes (clicks) of an optical experiment. The setup consists of entangled photon pairs, one of which is sent an interferometer emulating a finite temperature amplitude damping device. Blocking specific paths of the interferometer is tantamount to restricting the possible configurations of the reservoir. And by measuring its entangled pair, we can directly implement the two-point measurement scheme, thus avoiding the destructive nature of photo-detection.
\end{abstract}

\maketitle

%
%
{\bf \emph{Introduction --- }}
%
%
In the micro- and mesoscopic domains, thermodynamic quantities such as heat, work and entropy production may fluctuate significantly. 
They must therefore be described by random variables, with an associated probability distribution. 
This change in paradigm has led to ground-breaking new insights in  non-equilibrium thermodynamics. 
One reason, in particular, was the discovery of Fluctuation Theorems (FTs),  special symmetries of these distributions which generalize the second law of thermodynamics~\cite{Gallavotti1995,Evans1993,Crooks1998,Jarzynski1999a,Lebowitz1999}.
Thermodynamics, however, deals with processes, not states.
That is,  quantities such as heat or work depend on the \emph{transformations} a system undergoes.  
Assessing these in an experiment therefore requires measuring the system in (at least) two instants of time. 
For classical processes, this is generally not an issue~\cite{Liphardt2002,Collin2005,Gomez-Solano2011}. 
But in the quantum domain, measurements become invasive.

\begin{figure*}
\centering
\includegraphics[width=\textwidth]{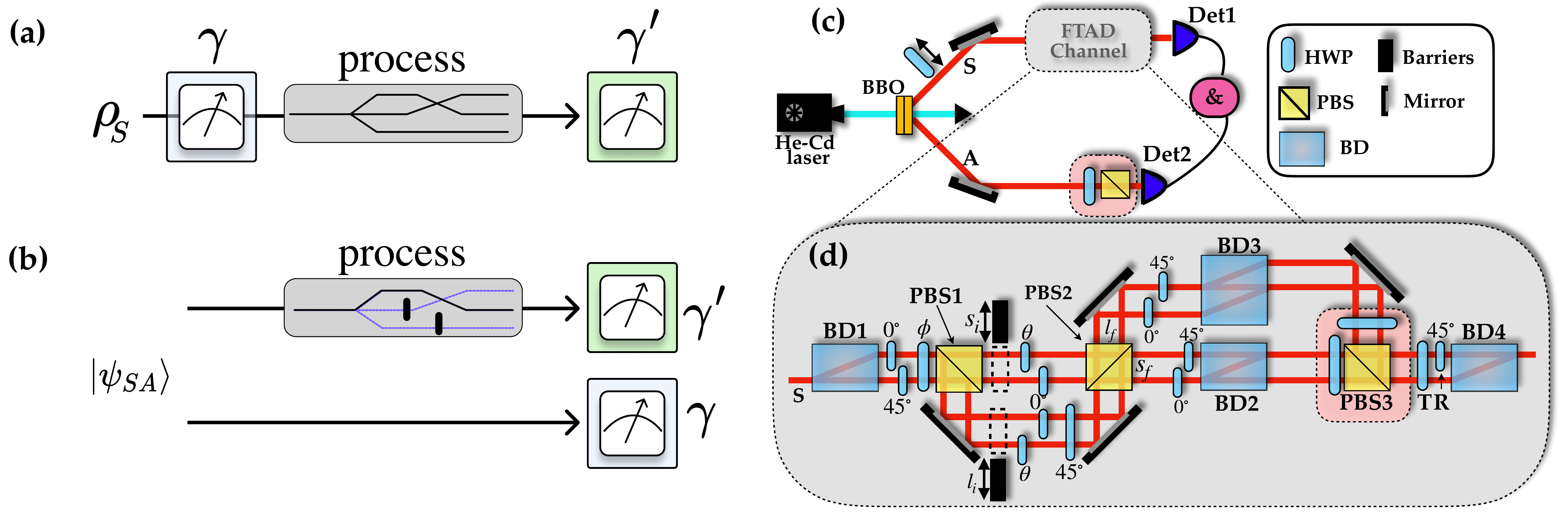}
\caption{
(a) The standard TPM scheme, which consists in measuring a quantum system before and after a thermodynamic process. 
The first measurement, with outcome $\gamma$, may invasively affect the process and, in general, is difficult to implement experimentally in a single experimental run.
(b) We bypass these issues using two features. First, we  distinguish the different configurations of the bath (black and purple lines inside the gray rectangle) by blocking specific arms of the interferometer. 
And second, we use entangled photon pairs, where only one is sent through the thermodynamic process. 
Due to the  coincidences postselection, measuring the other photon yields information about the initial state, before the process. 
(c) Experimental setup.  Pairs of entangled photons are created using SPDC. Photon in mode S is sent to a FTAD channel, while photon in mode A is detected after a projective measurement in its polarization. Coincidence counts are registered with detections in Det1 and Det2. 
(d) Interferometric implementation of the FTAD channel (see main text and appendix for more details).
\label{fig:drawing}
}
\end{figure*}

The default protocol for estimating quantum thermodynamic quantities is the two-point measurement (TPM) scheme~\cite{Talkner2007,Esposito2009,Campisi2011}  (Fig.~\ref{fig:drawing}(a)). 
It consists in measuring the system in the energy eigenbasis, before and after the process. 
This yields a stochastic trajectory of outcomes, $\gamma \to \gamma'$,  from which the corresponding thermodynamic quantities can be computed. 
However, experimentally implementing the two measurements \emph{in sequence} can be extremely challenging: Quantum observables  are often only inferred indirectly or via a destructive process, such as photo-detection. 

Instead, the distribution of outcomes $P(\gamma,\gamma')$ is usually written as $P(\gamma,\gamma') = P(\gamma'|\gamma) P(\gamma)$, and the two terms are determined from separate experiments. 
This was the case, for instance, in  Ref.~\cite{An2014}, which was the first to report on quantum fluctuation theorems.
In one experiment the system is prepared in some initial state (usually thermal), and the initial  probabilities $P(\gamma)$ are measured. 
Then, in a second experiment one \emph{prepares} a given state $\gamma$,  run the process and then measure it in $\gamma'$, thus obtaining the transition probability $P(\gamma'|\gamma)$. 
The TPM is then reconstructed from post-processing, with the purpose of confirming the validity of FTs. 
Similar issues are also found in other platforms~\cite{Hernandez-Gomez2020,Medeiros2018,Talarico2016}.

In the last years, there has been significant progress in developing alternative implementations of the TPM. 
One approach, used e.g. in nuclear magnetic resonance experiments~\cite{Batalhao2014,*Batalhao2015,Camati2016}, is based on Ramsay interferometry to indirectly estimate the characteristic function (from which the distribution can be reconstructed via a Fourier transform)~\cite{Mazzola2013a,Dorner2013}.
Another method is based on the interpretation of work  as the direct outcomes of a generalized quantum measurement (POVM), as put forth in~\cite{Roncaglia2014a}. This was implemented in Ref.~\cite{Cerisola2017}, to directly measure the work distribution in a quantum gas, without having to resort first to the TPM. 
Continuous weak measurements have also been used in superconducting circuits~\cite{Naghiloo2017,*Naghiloo2018}, as an alternative to the two-point nature of the TPM. 
To the best of our knowledge, the only experiment directly performing the two measurements of the TPM in the same experiment is the one in Ref.~\cite{Masuyama2017}, where nondemolition measurements on superconducting qubits were employed. 

In this letter we report  an experiment where both outcomes of the TPM are directly associated with the clicks of a single experiment. 
We study the entropy production in a system comprising an entangled photon pairs, where one of the photons is sent to an interferometer implementing a finite temperature amplitude damping (FTAD) channel, representing the interaction with a heat bath~\cite{Jarzynski2004a,Jevtic2015a,Manzano2017a,Santos2019}. 
To perform the TPM, we introduce two features, which are unique of our setup (Fig.~\ref{fig:drawing}(b)). 
First, each path of the interferometer is mapped into a different bath configuration. 
Hence, by blocking all but one of the paths, we can measure the statistics conditioned on a given bath configuration. 
Second, we use coincidence counts for measuring both photons in the entangled pair. 
Detecting the polarization of the photon that went through the interferometer yields $\gamma'$. 
And the outcomes of the entangled pair determines the configuration $\gamma$, before the process occurred. 
An experiment similar in spirit, but focused on a work protocol, has just been reported in~\cite{Solfanelli2021}.


%
%
{\bf \emph{Experimental setup --- }}
%
%
The experimental setup is shown in Fig. \ref{fig:drawing}(c).  
A He-Cd laser pumps two beta-barium-borate (BBO) crystals in a crossed axis configuration. By type-I spontaneous parametric down-conversion (SPDC), photon-pairs  in the entangled state $|\psi_{SA}\rangle = \sqrt{\delta} |00\rangle_{SA} + \sqrt{1-\delta} |11\rangle_{SA}$ are produced in the modes S and A \cite{kwiat99}, where the  computational basis  $|0\rangle_{S/A}$ and $|1\rangle_{S/A}$ corresponds to the horizontal and vertical polarizations, respectively.
The reduced state of S thus reads $\rho_S = \delta |0\rangle_S\langle 0| + (1-\delta) |1\rangle_S\langle 1|$,  which is a thermal state, with occupation $\delta$. 

The A-photons are detected  in an avalanche photon-diode  (Det2), after polarization projective measurements, implemented by a half-wave plate (HWP) and a polarized beam-splitter (PBS).
Due to the correlated nature of $|\psi_{SA}\rangle$ and the postselection of the detection of coincidences, outcomes 0,1 for A  imply initial states $\gamma = 0,1$ for S, and occur with probability 
$p_\gamma = \{\delta, 1-\delta\}$. 

The S-photons, on the other hand, pass through an  interferometer which implements an FTAD channel 
of the form $\Lambda[\rho_S] = \sum_{j=1}^4 E_j \rho_S E_j^\dagger$, with Kraus operators (satisfying $\sum_j E_j^\dagger E_j = \mathbb{1}$)
\begin{align}
E_{1}=\sqrt{p}\left[\begin{array}{cc}
1 & 0\\
0 & \sqrt{\eta}
\end{array}\right]\qquad & E_{2}=\sqrt{p}\left[\begin{array}{cc}
0 & \sqrt{1-\eta}\\
0 & 0
\end{array}\right]\label{eq:GADKraus}\\
E_{3}=\sqrt{1-p}\left[\begin{array}{cc}
\sqrt{\eta} & 0\\
0 & 1
\end{array}\right]\qquad & E_{4}=\sqrt{1-p}\left[\begin{array}{cc}
0 & 0\\
\sqrt{1-\eta} & 0
\end{array}\right],\nonumber 
\end{align}
which can be thought  of as a combination of emission ($E_{1,2}$) and excitation ($E_{3,4}$) processes with coupling strength $\eta \in [0,1]$ and  probabilities $p$ and $1-p$~\cite{chuang00}.  
The FTAD mimics a thermal bath acting on the photon, with the parameter $p$ interpreted as the thermal occupation probability. 
In fact, as discussed in~\cite{SupMat}, each index $j=1,2,3,4$ can also be associated to a pair of initial and final states of the reservoir. 
%

\begin{figure*}
    \centering
    \includegraphics[width=\textwidth]{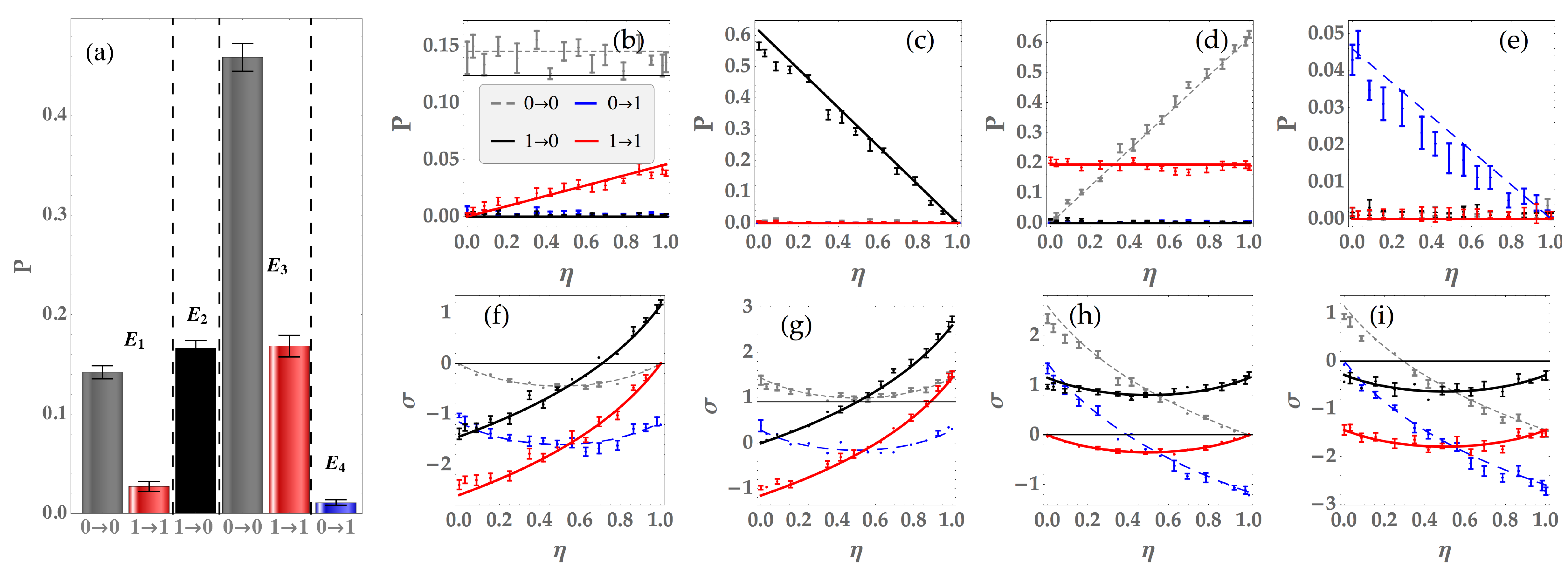}
    \caption{Experimentally determined probabilities and entropy production for all possible stochastic trajectories. (a)  Probabilities~\eqref{prob} for Kraus channels $E_1,\ldots, E_4$ and the system transitions $\gamma \to \gamma'$, with $\eta = 0.7(2)$.
    (b)-(e) Same, but as a function of $\eta$. Each  corresponds to a different channel $E_i$. 
    Each curve refers to a transition $\gamma\to\gamma'$, as labeled in image (b). 
    (f)-(i) Corresponding stochastic entropy production $\sigma[\gamma, \gamma', j]$, Eq.~\eqref{sigma}. 
    Points correspond to the experimental data, while the curves refer to the theoretical predictions. 
    Error bars are calculated from a Poissonian statistics for each outcome.
    In all curves $p = 0.19(1)$ and $\delta = 0.77(1)$.}
    
    \label{fig:data}
\end{figure*}

The  FTAD interferometer is implemented by entangling the photon polarization with path degrees of freedom,  as sketched in Fig.~\ref{fig:drawing}(d). The S-photons are first sent to a birrefringent calcite beam displacer BD1 that transmits (deviates) the vertically (horizontally) polarized photons, creating two  spatial modes (up ``$u$" and down ``$d$"), which we will call transversal modes (TMs). 
The polarization of photons in mode $d$  is rotated by a HWP at 45$^\circ$ and both TMs pass through a HWP, whose axis is rotated by $\phi/2$.  
The photons are thus reflected (transmitted) in PBS$_1$ with probability $p=\sin^2\phi$ ($1-p=\cos^2\phi$) into the longitudinal mode (LM)  long   ``$l_i$" (short ``$s_i$"), where the set of Kraus operators $\{E_1,E_2\}$ ($\{E_3,E_4\}$) are implemented.
In our setup, we tune the coupling strength of the channel with the angle $\theta/2$ of two HWPs, such that  $\eta=\cos^2\theta$. Hence, both parameters of the FTAD are adjusted at will. 
Finally, the photons are sent to PBS2, which incoherently combines both LMs, and  splits the photons in the two final states of the reservoir, $s_f$ and $l_f$. All HPWs in 0$^\circ$ serve to compensate the path length, allowing a coherent superposition of the TMs at BD2 and BD3, necessary for the implementation $E_1$ and $E_3$. BD4 plays two roles in our setup: (i) it traces out the TM and, in conjunction with the  plate ``T'', (ii) it  selects photons coming from $l_f$ or $s_f$, performing a projective measurement in the computational basis of the reservoir. The optical elements in the pink box are responsible for performing projective polarization measurements, yielding final outcomes $\gamma' = 0,1$. The joint distribution $P(\gamma,\gamma')$ is then obtained by registering coincidence counts between Det1 and Det2.

As discussed above, and formally demonstrated in the supplemental material \cite{SupMat}, each Kraus operator $E_j$ is associated to a specific optical path. 
Crucially, by blocking 3 out of the 4 paths, we can thus also determine $P(\gamma,\gamma',j)$, which represents the joint probability that the system undergoes a transition from $\gamma \to \gamma'$ in path $j=1,2,3,4$, which is given by 
\begin{equation}\label{prob}
P(\gamma,\gamma',j) = p_{\gamma} |\langle \gamma' | E_j |\gamma\rangle|^2.
\end{equation}
Thermodynamically, this would be tantamount to the joint TPM distribution of both system and bath, where the index $j$ collectively describes the initial and final state of the bath. 
In our case, this bath is represented instead by the path degrees of freedom of the interferometer.

The experimentally obtained probabilities are shown in Fig.~\ref{fig:data} for fixed $p = 0.19(1)$ and $\delta= 0.77(1)$.
Image (a) summarizes all probabilities which are non-zero, with fixed $\eta = 0.7(2)$. 
In fact, out of the possible $16$ outcomes $(\gamma,\gamma',j)$, only 6 are not zero. 
This is because $E_1$ and $E_3$ do not generate jumps, and hence allow only for the transitions $0\to 0$ and $1\to 1$ in the system. 
Conversely, $E_2$ and $E_4$ necessarily cause the system to jump, from $1\to 0$ and from $0\to 1$, respectively. 
This  is a manifestation of the preservation of the total number of excitations in system and bath.
Figs.~\ref{fig:data}(b)-(e) depicts the probabilities as a function of $\eta$, each plot corresponding to a different Kraus channel $E_j$.  
For comparison, the theoretical predictions from Eq.~\eqref{prob} are also shown.
In all cases, the fidelity  (Bhattacharyya distance) between theory and experiment was above 0.98, therefore confirming that the TPM protocol was successfully implemented.

%
%
{\bf \emph{Stochastic entropy production --- }}
%
%
The stochastic entropy production along a trajectory $(\gamma,\gamma',j)$ reads~\cite{Breuer2003,Manzano2017a,Landi2020a} 
\begin{equation}\label{sigma}
\sigma(\gamma,\gamma', j) = \ln(p_\gamma/\tilde{p}_{\gamma'}) + \Phi_j,
\end{equation}
where $\tilde{p}_{\gamma'} = \sum_{\gamma,j} P(\gamma,\gamma',j)$ is the probability associated to the final state of S. 
The first term in~\eqref{sigma} represents the stochastic variation in the system's entropy. 
Indeed, averaging over Eq.~\eqref{prob} leads to $\langle \ln p_\gamma/\tilde{p}_{\gamma'}\rangle = S(\Lambda[\rho_S]) - S(\rho_S)$, where $S(\rho) = - \tr(\rho\ln \rho)$ is the von Neumann entropy. 
The second term in Eq.~\eqref{sigma} is the entropy flux $\Phi_j$ to the  FTAD interferometer. 
It therefore depends only on the path $j$ the photon undergoes (and not on $\gamma,\gamma'$).
These fluxes are associated to quantum jumps in the system~\cite{Breuer2003}. 
Since $E_1$ and $E_3$ involve no jumps, we have  $\Phi_1 = \Phi_3 = 0$. 
Conversely, paths $2$ and $4$  must be accompanied by finite fluxes $\Phi_2$ and $\Phi_4$, since they involve jumps from  $1\to0$ and $0\to 1$, respectively.
To determine $\Phi_2$ and $\Phi_4$ we use the fact that the time-reversed Kraus operators $\tilde{E}_j$ should be related to the forward Kraus operators according to $\tilde{E}_j = e^{-\Phi_j/2} E_j^\dagger$~\cite{Crooks2008b,Manzano2017a}.
For the FTAD this yields the unique solution $\Phi_2 = - \Phi_4 = \ln p/(1-p)$.  
With these expressions for $\Phi_j$, it then follows from Eq.~\eqref{sigma} that $\sigma$ satisfies  the integral fluctuation theorem
\begin{equation}\label{FT}
\langle e^{-\sigma}\rangle = 1.
\end{equation}

The experimentally determined values of the stochastic entropy production~\eqref{sigma} are shown in Fig.~\ref{fig:data}(f)-(i) as a function of $\eta$ for fixed $p = 0.19(1)$ and $\delta= 0.77(1)$. 
Each plot corresponds to a different channel $j$ of the FTAD. 
For some trajectories the corresponding probabilities (Fig.~\ref{fig:data}(b)-(e)) are identically zero, so that these values will not contribute to any averages. 
For instance, because of Fig.~\ref{fig:data}(b), in Fig.~\ref{fig:data}(f) the only values of $\sigma$ which will actually play a non-trivial role are 
those corresponding to the trajectories $0\to 0$ and $1\to 1$ (gray and red curves).
Experimentally, we can still determine $\sigma(\gamma,\gamma', j)$ even when $P(\gamma,\gamma',j) \equiv 0$. 
This follows from Eq.~\eqref{sigma}, since the probability $\tilde{p}_{\gamma'}$ is an average over all trajectories. 

\begin{figure}[h!]
    \centering
    \includegraphics[width=8.5cm]{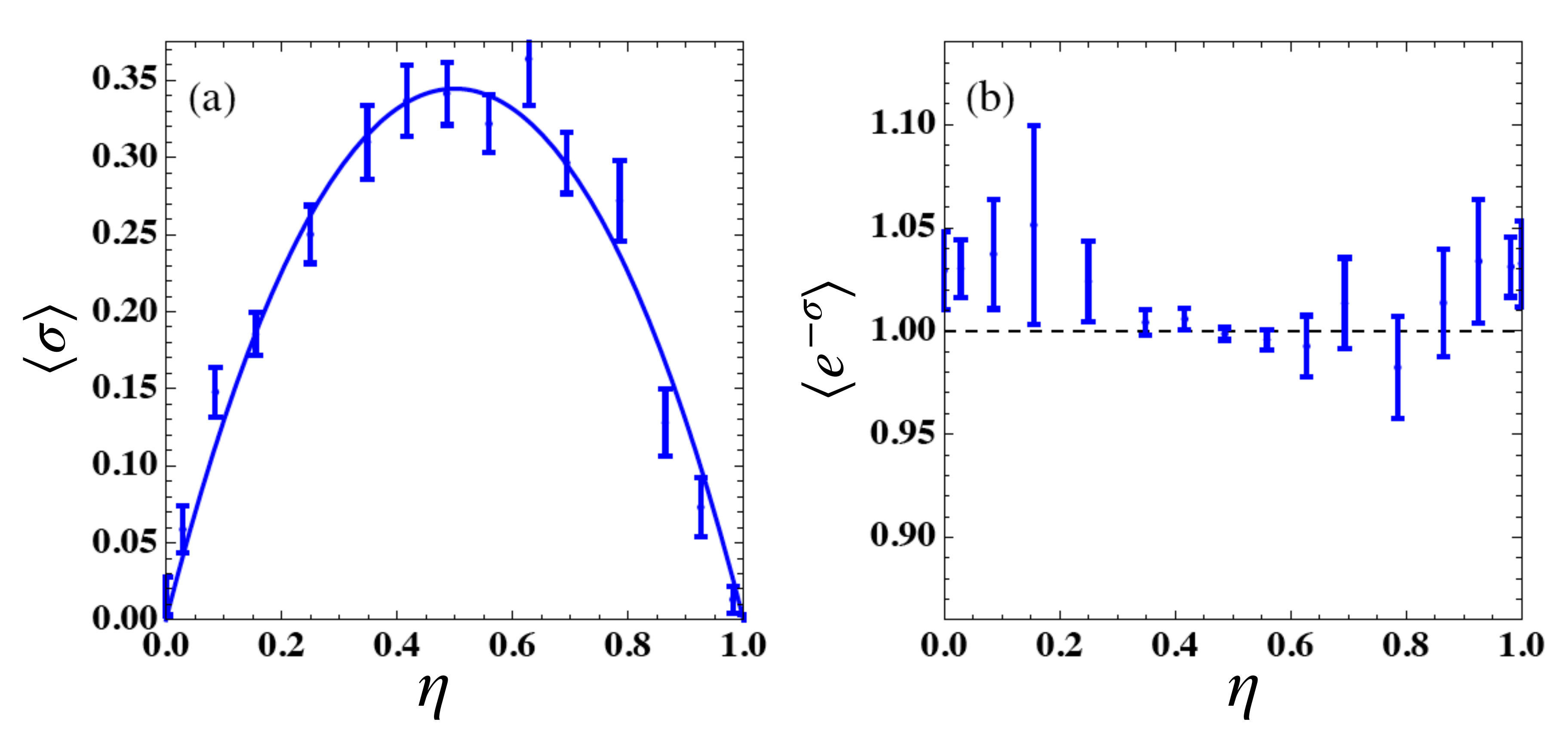}
    \caption{
    (a) Average entropy production $\langle \sigma \rangle$ and 
    (b) fluctuation theorem $\langle e^{-\sigma}\rangle$ as a function fo $\eta$, for $p = 0.19(1)$. }
    \label{fig:mean_entropy_production}
\end{figure}

As can be seen in Fig.~\ref{fig:data}(f)-(i), at the stochastic level some of the entropy productions can be negative. 
This does not contradict the second law, which holds only at the level of averages. 
Indeed, from Eq.~\eqref{FT} and Jensen's inequality, it follows that $\langle \sigma \rangle \geqslant 0$. 
This means that negative values are less likely, which can be observed by comparing the probabilities in Fig.~\ref{fig:data}(b)-(e).
The average entropy production $\langle \sigma\rangle = \sum_{\gamma,\gamma', j} \sigma(\gamma,\gamma',j) P(\gamma,\gamma',j)$ is shown in Fig.~\ref{fig:mean_entropy_production}(a), together with the fluctuation theorem~\eqref{FT} in Fig.~\ref{fig:mean_entropy_production}(b).
In both cases, the experiment confirms the experimental predictions: $\langle \sigma \rangle \geqslant 0$ and $\langle e^{-\sigma}\rangle = 1$.


%


{\bf \emph{Significance - }} The key feature of our setup is the ability to obtain the full TPM statistics directly from the clicks of an experiment. 
This was accomplished by combining two features. 
First, the configurations of the reservoir are characterized by a set of Kraus operators, each corresponding to a different optical path (see the appendix).
Hence, by blocking 3 out of the 4 paths, we can directly study the statistics conditioned on each bath configuration. 
Second, we use an entangled photon pair to non-destructively measure the initial system configuration. 
Combined with the final measurement, associated to the photon that went through the interferometer, this yields the full TPM statistics. 

Thermodynamic quantities, such as heat, work or entropy production, characterize transformations (i.e. processes) that the system undergoes. 
But directly assessing processes in quantum systems is notoriously difficult. 
Our setup offers a platform for overcoming this. 

As a proof-of-principle, we have focused on the entropy production in a heat exchange process. 
This is convenient since entropy production is a fully information-theoretic quantity, and hence avoids entering into issues about the energetics of photon processes. 
However, our approach is highly flexible and can be extended to various other thermodynamic protocols, including systems with initial coherences. 
The latter, in particular, would be an interesting direction of future research, since the TPM  becomes invasive in the presence of energetic coherences~\cite{Aberg2016a,Perarnau-Llobet2017,Levy2019,Guerardini2020,Micadei2019,Deffner2016,Sone2020a}.
Our framework could then be used to compare the TPM with other alternatives that have recently been proposed to overcome this invasiveness, such as  Bayesian networks~\cite{Micadei2019} and quasiprobabilities~\cite{Levy2019,YungerHalpern2018a}.  

\begin{acknowledgements}
We thank Stephen Walborn for valuable discussions. The authors acknowledge financial support from the Brazilian agencies CNPq (PQ Grants No. 307058/2017-4m No.   and INCT-IQ)
GTL acknowledge the financial support of the Sao Paulo Funding Agency FAPESP 
(Grants No. 2017 /50304-7, 2017/07973-5 and 2018/12813-0).
The authors acknowledge the financial support of the Brazilian funding agency CNPq (PQ Grants No. 305740/2016-4, 307058/2017-4, 309020/2020-4 and INCT-IQ 246569/2014-0).
TLS would like to thank Serrapilheira Institute (Grant No. Serra-1709-17173).  
This work was realized as part of the CAPES/PROCAD program.  

\end{acknowledgements}


%

\pagebreak
\widetext

\newpage 
\begin{center}
\vskip0.5cm
{\Large Supplemental Material}
\end{center}
\vskip0.4cm

\setcounter{section}{0}
\setcounter{equation}{0}
\setcounter{figure}{0}
\setcounter{table}{0}
\setcounter{page}{1}
\renewcommand{\theequation}{S\arabic{equation}}
\renewcommand{\thefigure}{S\arabic{figure}}

\vskip0.7cm

In this supplemental material we provide additional details on the implementation of the Finite Temperature Amplitude Damping (FTAD) channel used in the main text [Eq.~(1)].
We first discuss the connection between the channel and the interaction between a system and a thermal bath. 
Then we give additional details on how this FTAD emerges from the interferometer used in our experiment.

\section{FTAD and interaction with a thermal reservoir}
\label{sec:appendixA}

We consider a system $S$ interacting with an environment $E$,  prepared in an arbitrary uncorrelated state  $\rho_S \otimes \rho_E$.
The two are assumed to evolve unitarily according to 
\begin{equation}
\rho_{SE}^{\prime}= U\rho_{S}\otimes\rho_{E}U^{\dagger},
\label{eq:Init_state}
\end{equation}
The transformation over the system state is recovered by tracing out
the environmental degrees of freedom 
\begin{equation}
\rho^{\prime}_{S}=\sum_{k}\bra{k}U\rho_{S}\otimes\rho_{E}U^{\dagger}\ket{k},\label{gen_map}
\end{equation}
where $\{\ket{k}\}$ is an orthonormal basis of the environment  \cite{chuang00}. If $\rho_{E}$ is a pure state, say $\rho_{e}=\ketbra{0}{0}$, then
we can identify the Kraus operators with $E_{k}=\bra{k}U\ket{0}$. Thus,
 the allowed number of independent Kraus operators is equal to the  dimension of
the  Hilbert space of the environment. For example, a FTAD channel
would require a four-dimensional environment. Instead, we can assume that the environment is initially in a mixed state, which allows us to mimic the channel using an environment  with dimension 2, which is half of the number of Kraus operators \cite{Horodecki1999}. It is not always possible to reconstruct a channel in this fashion \cite{Terhal1999},
but particularly for the FTAD it is.

Thinking about the physical interpretation of the FTAD channel as an interaction of a system with a thermal bath, it
makes sense to consider  the initial state of the environment as a thermal state $\rho_{E}=p\ketbra{0}{0}+(1-p)\ketbra{1}{1}$, whose temperature
is determined by the channel parameter $p$. Plugging this into equation (\ref{gen_map}) leads to 
\begin{equation}
\rho^{\prime}_{S}=\sum_{k=0,1}\left[\sqrt{p}\bra{k}U\ket{0}\right]\rho_{S}\left[\sqrt{p}\bra{0}U^{\dagger}\ket{k}\right]+
 \left[\sqrt{1-p}\bra{k}U\ket{1}\right]\rho_{S}\left[\sqrt{1-p}\bra{1}U^{\dagger}\ket{k}\right].
\end{equation}
This is a completely positive and trace preserving map with Kraus operators 
\begin{equation}
\begin{split}
E_{1}=\sqrt{p}\bra{0}U\ket{0}&\qquad E_{2}=\sqrt{p}\bra{1}U\ket{0}\\
E_{3}=\sqrt{1-p}\bra{1}U\ket{1}&\qquad E_{4}=\sqrt{1-p}\bra{0}U\ket{1}.\label{UKraus}
\end{split}
\end{equation}
One can see that  the Kraus operators in Eq. (\ref{eq:GADKraus}) are recovered by considering the following two-qubit unitary transformation 
\begin{equation}
U=\left[\begin{array}{cccc}
1 & 0 & 0 & 0\\
0 & \sqrt{\eta} & -\sqrt{1-\eta} & 0\\
0 & \sqrt{1-\eta} & \sqrt{\eta} & 0\\
0 & 0 & 0 & 1
\end{array}\right],
\end{equation}
 where the first and the second qubit are the system and the reservoir, respectively. This is a partial swap unitary, which preserves the number of excitations in the global system and the coupling $\eta$ gives the probability that the system decays, emitting one excitation to the reservoir.


\section{Detailed explanation of the experimental setup }
\label{sec:appendix}

\begin{figure}
\centering
\includegraphics[width=0.5\textwidth]{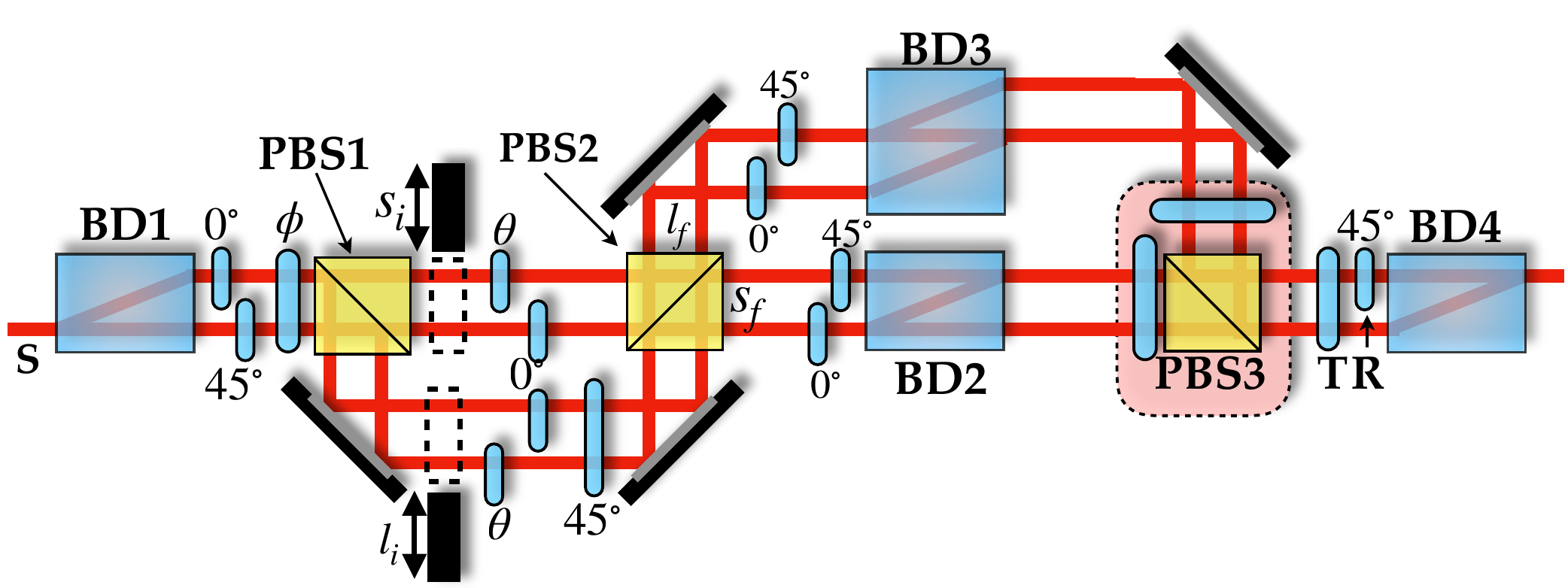}
\caption{\label{fig:drawing_sup_mat} Interferometer used in our experiment to implement the FTAD channel.}
\end{figure}

In this appendix we describe the net transformation of the interferometer shown in Fig. \ref{fig:drawing}(d), and reproduced  in Fig.~\ref{fig:drawing_sup_mat}. 
We will also demonstrate that  each path of the photons  corresponds to a trajectory where only one of the Kraus operators is applied. For this,  let us consider, without any loss of generality, that the polarization of the S-photons is prepared in a pure state $\ket{\psi_0}=a\ket{0}+b\ket{1}$.  After passing through the  BD1, this state is transformed  into 
\begin{equation}
\ket{\psi_1}= a\ket{0}\ket{u}+b\ket{1}\ket{d},
\end{equation}
where $\ket{u}$ and $\ket{d}$ are the TM up and down, respectively. Then, the polarization of photons in mode $d$  is rotated by a HWP at 45$^\circ$. This is equivalent to a Pauli $X$-gate in the polarization qubit controlled by the TM qubit, which  transfers the polarization state to the TM state, resulting in
 \begin{equation}
\ket{\psi_2}=\ket{0}\left(a\ket{u}+b\ket{d}\right).
\end{equation}
 Both TMs pass through a HWP, whose axis is rotated by an angle $\phi/2$, transforming the photon's state to 
 \begin{equation}
\ket{\psi_3}=(\cos{\phi}\ket{0}+\sin{\phi}\ket{1})\left(a\ket{u}+b\ket{d}\right).
\end{equation}
Afterwards, the  PBS1 creates two new paths, the longitudinal short ``$s_i$" and long ``$l_i$" paths, by transmitting horizontally
polarized photons while reflecting the vertically polarized. The state of the photons can be written as:
\begin{equation}
\ket{\psi_4}=\left(\cos{\phi}\ket{0}\ket{s_i}+\sin{\phi}\ket{1}\ket{l_i}\right)\left(a\ket{u}+b\ket{d}\right)
\end{equation}
We now identify, within the TM and LM,  which are the modes that actually constitute the degrees of freedom of the environment.  By tracing out the polarization,  we obtain a thermal state in  LM, whose temperature is controlled by the angle $\phi$, multiplied by the state $\ket{\psi_0}$, transferred to  the TM. This is exactly the initial product state in Eq. (\ref{eq:Init_state}), meaning that the initial excited and ground state of the reservoir are related with the longitudinal modes $s_i$ and $l_i$, respectively.  As mentioned in the main text, the photons will go through the $l_i$ or the $s_i$ path with probabilities $p=\sin^2\phi$ and $1-p=\cos^2\phi$. Given that the coherence length of the SPDC photons is much smaller than the size
of the path difference between $s_i$ and $l_i$, then there is no coherence between  $s_i$ and $l_i$ at PBS2. The same happens after PBS2 when the photons can follow two final LM $s_f$ and $l_f$ with different lengths, being incorehently recombined in PBS3. In this last case, each path  is associated with one of the computational basis states of the bath after the interaction with the system. Therefore,  after tracing out the corresponding path, each combination of $\{s_i,l_i\}$ and $\{s_f,l_f\}$ paths performs the transformation of one of the Kraus operators composing the FTAD channel. In what follows, we demonstrate this  by considering each path separately. It is worth noting that we will write sub-normalized states in each step such that all the parameters of the Kraus operator appear. The correct normalizations comes when considering the total transformation. 

\textbf{Long $l_i$ path.} The unnormalized state of the photons that are reflected by PBS1 can be written as $\ket{\psi_{4l_i}}=\sin\phi\ket{1}(a\ket{u}+b\ket{d})$. This is transformed by a set of HWPs into the state
\begin{equation}
 \ket{\psi_{5l_i}}=\sin\phi\left[-a\ket{0}\ket{u}+b\left(-\cos\theta\ket{0}+\sin\theta\ket{1}\right)\ket{d}\right].
\end{equation}
Given that the coupling strength of the channel is related with the angle $\theta$, one can notice that only the photons in  $d$ interact with the bath, which is related to the initial state $\ket{1}$ of the polarization (see $\ket{\psi_0}$). Provided that the FTAD is an operation that preserves the number of excitations, the  longitudinal mode $l_i$ represents the initial state $\ket{0}$ of the reservoir.  

After reaching PBS2, the photons can follow $s_f$ or $l_f$, depending on its polarization. Its final state  in each path after BD2 and BD3 reads:
\begin{itemize}
 \item Short $s_f$ path:
 \begin{equation}\label{eq:ls}
  \ket{\psi_{6l_is_f}}=\sin\phi\left(b\sin\theta\ket{1}\right)\ket{d}\ket{s_f}
 \end{equation}
 \item Long $l_f$ path:
 \begin{equation}\label{eq:ll}
  \ket{\psi_{6l_il_f}}=-\sin\phi\left(a\ket{1}+b\cos\theta\ket{0}\right)\ket{d}\ket{l_f}.
 \end{equation}
\end{itemize}
One can see that the role of BD2 and BD3 along with the preceding HWPs is to transfer back the state from the TMs to the polarization. The TMs are just  auxiliary degrees of freedom used to properly implement the FTAD in the polarization of the photons.

\textbf{Short $s_i$ path.} The photons transmitted in PBS1 are in a unnormalized state $\ket{\psi_{4s_i}}=\cos\phi\ket{0}(a\ket{u}+b\ket{d})$, which is transformed by the set of HWPs into
\begin{equation}
 \ket{\psi_{5s_i}}=\cos\phi\left[a\left(\cos\theta\ket{0}+\sin\theta\ket{1}\right)\ket{u}+b\ket{0}\ket{d}\right].
\end{equation}
Akin to what happens in the $l_i$ path, here the bath interacts only with  photons in mode $u$, which is related with the initial state $\ket{0}$ of the polarization. This means that the $s_i$ mode is associated with the initial state $\ket{1}$ of the reservoir. Again, we can write the photon state in each path $s_f$ and $l_f$  after BD2 and BD3 as:
\begin{itemize}
 \item Short $s_f$ path:
 \begin{equation}\label{eq:ss}
  \ket{\psi_{6s_is_f}}=\cos\phi\left(a\cos\theta\ket{1}+b\ket{0}\right)\ket{u}\ket{s_f}
 \end{equation}
 \item Long $l_f$ path:
 \begin{equation}\label{eq:sl}
  \ket{\psi_{6s_il_f}}=\cos\phi\left(a\sin\theta\ket{0}\right)\ket{u}\ket{l_f}.
 \end{equation}
\end{itemize}

The very final polarization state obtained after recombining incoherently all the longitudinal and transversal modes is given as 
\begin{equation}\label{eq:final}    
 \rho_f=\Tr_{TM,LM}\left[\ketbra{\psi_{6s_is_f}}{\psi_{6s_is_f}}+\ketbra{\psi_{6s_il_f}}{\psi_{6s_il_f}}
+\ketbra{\psi_{6l_is_f}}{\psi_{6l_is_f}}+\ketbra{\psi_{6l_il_f}}{\psi_{6l_il_f}}\right].                                                                                                                                                                                                                        
\end{equation}
Using Eqs. \eqref{eq:GADKraus}, \eqref{eq:ls}, \eqref{eq:ll}, \eqref{eq:ss}, and \eqref{eq:sl}, and taking  $p=\sin^2\phi$ and $\eta=\cos^2 \theta$, one can see that the trajectories $l_i l_f$, $l_i s_f$, $s_i s_f$ and $s_i l_f$ can be  identified with the terms $E_1\ketbra{\psi_0}{\psi_0}E_1^\dagger$, $E_2\ketbra{\psi_0}{\psi_0}E_2^\dagger$, $E_3\ketbra{\psi_0}{\psi_0}E_3^\dagger$, and $E_4\ketbra{\psi_0}{\psi_0}E_4^\dagger$, respectively, up to a Pauli-X gate interchanging the roles of $\ket{0}$ and $\ket{1}$.

Instead of recombining all modes together and then performing polarization measurements, we place a  HWP in the end of each $s_f$ and $l_f$ modes, which together with PBS3,  performs any desired projection onto the polarization.  BD4 and the wave plates "T" and "R" (fixed at 45$^{\circ}$) plays two roles in our setup: (i) to implement an incoherent combination of the TM modes, which is equivalent to tracing out this degree of freedom, and (ii) by mapping the information of the LM $s_f$ and $l_f$ before PBS3 to polarization after this optical element,  projective measurement onto the final $s_f$ and $l_f$ are performend by simply changing the angle of the plate T.

\end{document}